\documentclass[slac_one]{revtex4}
\usepackage{graphicx}
\usepackage{fancyhdr}
\newcommand{\ppbar}{\ensuremath{p\bar{p}}}
\newcommand{\ttbar}{\ensuremath{t\bar{t}}}

\newcommand{\JES}{\mathop{\mathrm{JES}}}
\newcommand{\DJES}{\Delta_{\JES}}

\begin{document}

\title{Top Quark Mass Measurement Using a Matrix Element Method with
  Quasi--Monte Carlo Integration} %% Paper title goes here


\author{Paul J. Lujan (for the CDF Collaboration)}
\affiliation{Lawrence Berkeley National Laboratory, Berkeley, CA
  94720, USA}

\begin{abstract}

We report an updated measurement of the top quark mass obtained from
$\ppbar$ collisions at $\sqrt{s} = 1.96$ TeV at the Fermilab Tevatron
using the CDF II detector. Our measurement uses a matrix element
integration method to obtain a signal likelihood, with a neural
network used to identify background events and a likelihood cut
applied to reduce the effect of badly reconstructed events. We use a
2.7 fb$^{-1}$ sample and observe 422 events passing all of our
cuts. We find $m_t$ = 172.2 $\pm$ 1.0 (stat.) $\pm$ 0.9 (JES) $\pm$ 1.0
(syst.) GeV/$c^2$, or $m_t$ = 172.2 $\pm$ 1.7 (total) GeV/$c^2$.

\end{abstract}

\maketitle

\thispagestyle{fancy}


\section{INTRODUCTION}

The top quark is the heaviest known particle in the Standard
Model. Its mass is an important parameter to be determined, both for
its intrinsic interest, and because precision measurements of the top
quark mass, in conjunction with the $W$ boson mass, allow us to set
constraints on the mass of the Higgs boson within the Standard
Model. In this letter we describe a precision measurement of the top
quark mass using a matrix element integration method. This measurement
uses 2.7 fb$^{-1}$ of data collected by the CDF II detector.

We obtain a top mass measurement by integrating over unmeasured
quantities in the matrix element using a quasi--Monte Carlo
integration. This allows us to minimize assumptions made about the
kinematics of an event, resulting in improved precision. The
integration method yields a likelihood curve as a function of the top
pole mass.

The largest source of systematic uncertainty in our measurement is the
jet energy scale (JES). To reduce our uncertainty due to this source,
we introduce an additional parameter to our likelihood, $\DJES$, which
allows us to use the information in the $W$ decay to determine the
JES. $\DJES$ parameterizes the shift in JES in units of the systematic
error for a given jet. Our likelihood is thus constructed as a
2D function of $m_t$ and $\DJES$; we then combine the likelihoods for
all events and eliminate $\DJES$ as a nuisance parameter to find a
final top mass value.

\section{EVENT SELECTION}

At the Fermilab Tevatron, top quarks are predominantly produced in
$\ttbar$ pairs, where the $t$ decays into a $W$ boson and a $b$ quark
$\sim$ 100\% of the time. The $W$ can then decay into a charged lepton
and a neutrino (``leptonic'' decay) or a quark-antiquark pair
(``hadronic'' decay). We search for events in the ``lepton + jets''
channel, where one $W$ decays hadronically and one leptonically. Thus,
we analyze events with four high-energy jets (two from the $b$ quarks
and two from the hadronic $W$ decay), at least one of which is
required to be $b$-tagged using a secondary vertex algorithm which
identifies tracks displaced from the primary vertex; exactly one high
energy electron or muon (from the leptonic $W$ decay); and large
missing transverse energy (from the neutrino).

The principal backgrounds to our signal are events where a $W$ boson
is produced in conjunction with heavy flavor jets ($b\bar{b}$,
$c\bar{c}$, or $c$), a $W$ boson is produced with light jets which are
mistagged as $b$-jets, and QCD events not containing a $W$ where the
$W$ signature is faked. Overall we expect 105.7 $\pm$ 42.0 background
events in our observed 494 candidate
events. Table~\ref{expected_backgrounds} shows our expected
backgrounds.

\begin{table}[t]
\begin{center}
\caption{Expected backgrounds for the $W$+4 tight jet sample used.}
\begin{tabular}{|l|c|c|}
\hline
Background                &  1 tag             & $\ge$ 2 tags    \\
\hline
non-W QCD                 & 20.0 $\pm$ 17.3    & 0.8 $\pm$ 1.6 \\
\hline
W+light mistag, diboson, or $Z$    & 27.7 $\pm$ 5.6     & 1.1 $\pm$ 0.1 \\
\hline
W+heavy ($b\bar{b}$, $c\bar{c}$, $c$) & 45.0 $\pm$ 37.8    & 6.0 $\pm$ 5.0 \\
Single top                & 3.8  $\pm$ 0.2     & 1.2 $\pm$ 0.2 \\
\hline
Total background          & 96.5 $\pm$ 36.8    & 9.2 $\pm$ 5.2 \\
Predicted top signal      & 259.4 $\pm$ 33.6   & 98.8 $\pm$ 16.0 \\
\hline
Events observed           & 389                & 105  \\ 
\hline
\end{tabular}
\label{expected_backgrounds}
\end{center}
\end{table}

% figures should be put into the text as floats.
% Use the graphics or graphicx packages (distributed with LaTeX2e)
% and the \includegraphics macro defined in those packages.
% See the LaTeX Graphics Companion by Michel Goosens, Sebastian Rahtz,
% entire page. There is no need to do explicit centering.

% \begin{figure}
% \end{figure}

% Surround figure environment with turnpage environment for landscape
% figure
% \begin{figure}
% \end{figure}

\section{MATRIX ELEMENT METHOD}

We calculate a two-dimensional likelihood as a function of $m_t$ and
$\DJES$ by integrating over the matrix element for $\ttbar$ production
and decay over the unknown parton-level quantities, using transfer
functions to connect these with the measured jets. Our overall
likelihood formula is:

\begin{equation}
  L(\vec{y} \mid m_t, \DJES) = \frac{1}{N(m_t)} \frac {1}{A(m_t,\DJES)} 
  \sum_{i\,=1}^{24} w_{i} L_{i}(\vec{y} \mid m_t, \DJES)
\end{equation}
with
\begin{equation}
   L_{i}(\vec{y} \mid m_t, \DJES) = \int \frac {f(z_1) f(z_2)}{FF}
	    ~\textrm{TF}(\vec{y}  \mid \vec{x}, \DJES)
           ~|M (m_t,\vec{x})|^{2} ~d\Phi(\vec{x}),
\end{equation}
\noindent
where $\vec{x}$ denotes the parton-level quantities, $\vec{y}$ denotes
the quantities measured in our detector, $M$ is the matrix element for
$\ttbar$ production and decay, $f(z)$ is the parton distribution
functions (PDFs) for the momenta of the two incoming particles, FF is
the flux factor normalizing the PDFs, $N(m_t)$ is a normalization
factor, $A(m_t, \DJES)$ is an acceptance factor to correct for the
effect of the event selection criteria, and $\Phi$ is the parton-level
phase space integrated over. The integral is evaluated for each of the
24 possible jet-parton assignments and then summed with appropriate
weights corresponding to the probability that a given jet-parton
assignment corresponds with the observed $b$-tags. We integrate over a
total of 19 variables. In order to perform this integral in a
practical amount of time, we employ quasi--Monte Carlo
integration~\cite{qmc}, which uses quasi-random sequences. These
sequences provide more uniform coverage of the phase space, resulting
in faster integral convergence than with normal Monte Carlo
techniques.

We use a neural network to identify events likely to be background,
and subtract out their contribution to the total likelihood by
estimating the average contribution for background events from Monte
Carlo. We also consider the effect of events which we call ``bad
signal''. These are events which contain an actual $\ttbar$ decay, but
where the final observed objects in our detector do not come directly
from $\ttbar$ decay (due to extra jets from initial or final state
radiation, $W \rightarrow \tau$ decay, or other causes). To reduce the
effect of these poorly-modeled events, we apply a cut of 10 to the
peak of the log-likelihood curve. In Monte Carlo simulation, this cut
eliminates 20\% of ``bad signal'' events and 27\% of background while
retaining 97\% of our good signal events.

\begin{figure*}[htb]
\centering
\includegraphics[width=.3\textwidth]{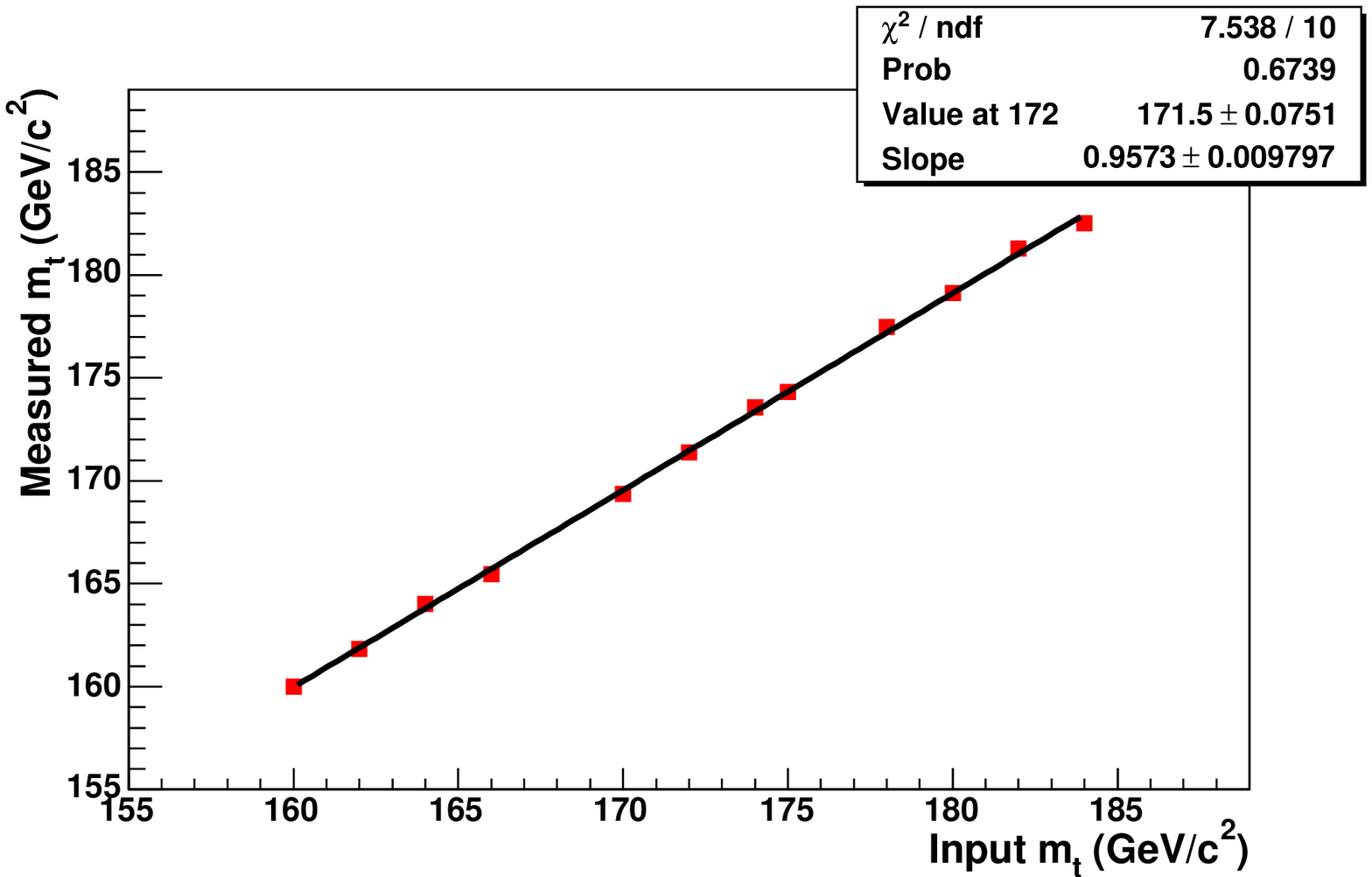}
\includegraphics[width=.3\textwidth]{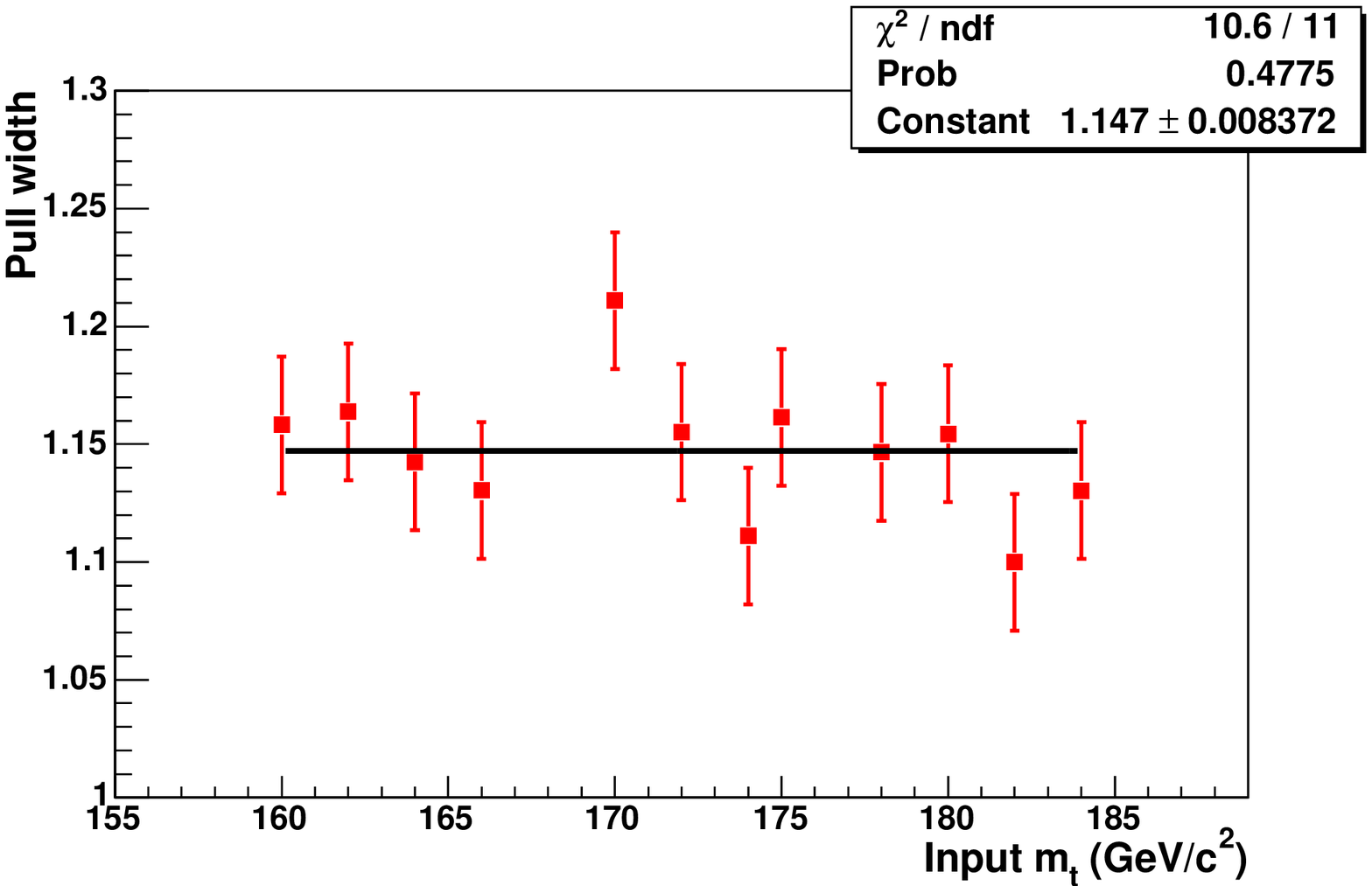}
\caption{Results using Monte Carlo events to test and calibrate our
  method. Left: measured mass vs. input mass. Right: pull width
  vs. input mass.}
\label{calib}
\end{figure*}

We test and calibrate our measurement using {\sc pythia} Monte Carlo
events over a variety of input $m_t$ and $\DJES$ values by performing
pseudo-experiments. Using the results of the pseudo-experiments, we
obtain a final set of calibration constants for our measured top mass
and statistical uncertainty. Figure~\ref{calib} shows the results of
our Monte Carlo testing. Figure~\ref{figures} shows the effect of the
likelihood cut on Monte Carlo events.

\section{RESULT}

We have 494 events passing our initial selection cuts, of which 422
events pass the likelihood cut as well. With these 422 events, we
measure:

\begin{center}
$m_t$ = 172.2 $\pm$ 1.0 (stat.) $\pm$ 0.9 (JES) $\pm$
1.0 (syst.) GeV/$c^2$ 
= 172.2 $\pm$ 1.7 (total) GeV/$c^2$
\end{center}

\begin{figure*}[b]
\centering
\includegraphics[width=.45\textwidth]{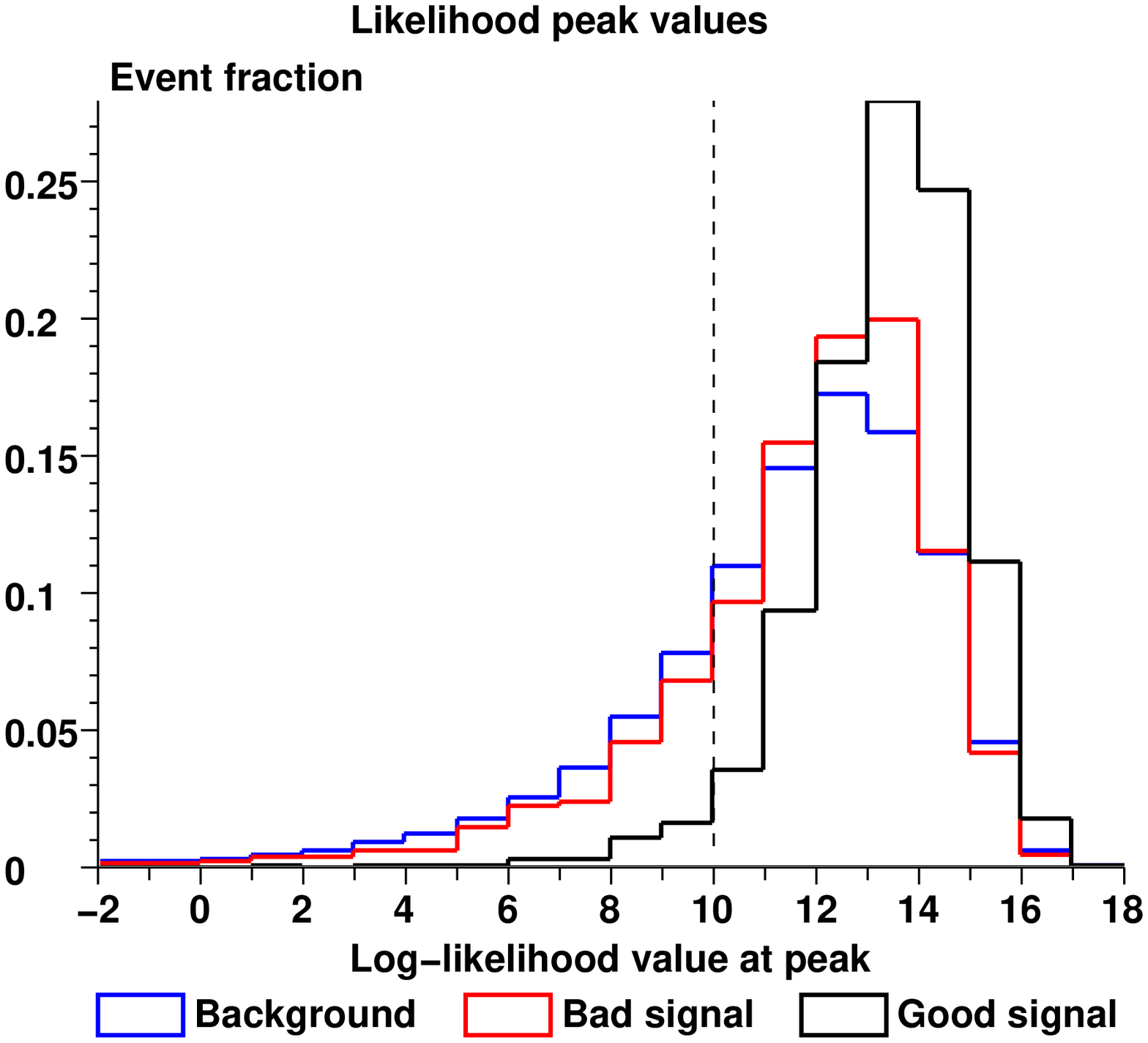}
\includegraphics[width=.52\textwidth]{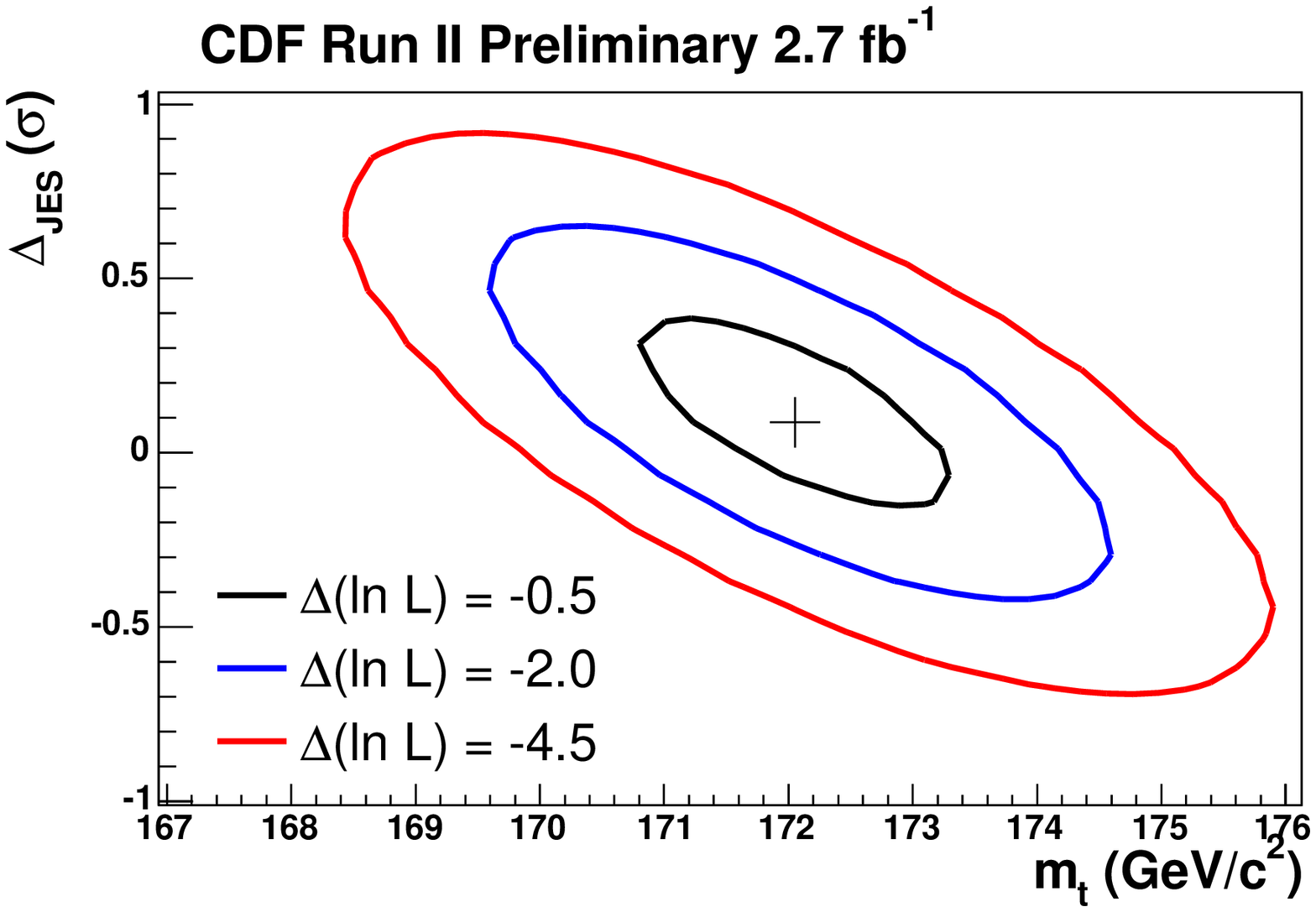}
\caption{Left: Value of the log-likelihood curve at its peak for good
  signal, bad signal, and background events in Monte Carlo. The dashed
  line shows the cut at 10 used. Right: Contours of the 2D likelihood
  distributions obtained with our final data sample. The contours
  shown correspond to a statistical uncertainty of 1, 2, and 3 $\sigma$.}
\label{figures}
\end{figure*}

Figure~\ref{figures} shows the final contours of 1-$\sigma$,
2-$\sigma$, and 3-$\sigma$ statistical uncertainty around the measured
value. The total result attains a precision of better than 1\% in
$m_t$.

Our main sources of systematic uncertainty are from the Monte Carlo
generator used for our calibration and testing (0.5 GeV/$c^2$), the
residual JES uncertainty resulting from variation of the individual
sources of our total JES uncertainty (0.5 GeV/$c^2$), uncertainty from
the modeling of the jet energy scale for $b$-jets (0.4 GeV/$c^2$), and
uncertainty in the background model (0.4 GeV/$c^2$). We also have
smaller uncertainties from initial-state and final-state radiation,
lepton $P_T$ measurement, pileup events, calibration, and PDFs (0.3,
0.2, 0.2, 0.1, and 0.1 GeV/$c^2$, respectively), for a total of 1.0
GeV/$c^2$.

%Table~\ref{systematics} summarizes the sources of systematic
% uncertainty in our measurement.

%\begin{table}[t]
%\caption{Systematic uncertainties for our measurement.}
%\label{systematics}
%\begin{center}
%\begin{tabular}{|r|c|}
%\hline
%Systematic source & Systematic uncertainty (GeV/$c^2$) \\
%\hline
%Calibration                & 0.1 \\
%MC generator               & 0.5 \\
%ISR and FSR                & 0.3 \\
%Residual JES               & 0.5 \\
%$b$-JES                    & 0.4 \\
%Lepton $P_T$               & 0.2 \\
%Pileup                     & 0.1 \\
%PDFs                       & 0.2 \\
%Background                 & 0.4 \\
%\hline
%Total                      & 1.0 \\
%\hline
%\end{tabular}
%\end{center}
%\end{table}


\begin{acknowledgments}
I would like to thank the funding institutions supporting the CDF
collaboration. A full list can be found in~\cite{mtm3_public}.
\end{acknowledgments}

\end{document}